# Selecting Source Code Generation Tools Based on Bandit Algorithms

Ryoto Shima
Kindai University
Higashi-osaka, Japan
2010370026k@kindai.ac.jp

Masateru Tsunoda
Kindai University
Higashi-osaka, Japan
tsunoda@info.kindai.ac.jp

Yukasa Murakami
Okayama University
Okayama, Japan
m.yukasa@gmail.com

Akito Monden
Okayama University
Okayama, Japan
monden@okayama-u.ac.jp

Amjed Tahir
Massey University
Palmerston North, NZ
a.tahir@massey.ac.nz

Kwabena Ebo Bennin
Wageningen UR
Wageningen, Netherlands
kwabena.bennin@wur.nl

Koji Toda
Fukuoka Institute of Tech.
Fukuoka, Japan
toda@fit.ac.jp

Keitaro Nakasai
OMU College of Tech.
Osaka, Japan
nakasai@omu.ac.jp

## ABSTRACT

**Background**: Recently, code generation tools such as ChatGPT have drawn attention to their performance. Generally, a prior analysis of their performance is needed to select new code-generation tools from a list of candidates. Without such analysis, there is a higher risk of selecting an ineffective tool, negatively affecting software development productivity. Additionally, conducting prior analysis of new code generation tools takes time and effort. **Aim**: To use a new code generation tool without prior analysis but with low risk, we propose to evaluate the new tools during software development (i.e., online optimization). **Method**: We apply the bandit algorithm (BA) approach to help select the best code-generation tool among candidates. Developers evaluate whether the result of the tool is correct or not. When code generation and evaluation are repeated, the evaluation results are saved. We utilize the stored evaluation results to select the best tool based on the BA approach. Our preliminary analysis evaluated five code generation tools with 164 code generation cases using BA. **Result**: The BA approach selected ChatGPT as the best tool as the evaluation proceeded, and during the evaluation, the average accuracy by the BA approach outperformed the second-best performing tool. Our results reveal the feasibility and effectiveness of BA in assisting the selection of best-performing code generation tools.

## 1 Introduction

Recently, code generation tools such as ChatGPT have attracted the attention of programmers [4] with their performance. The performance of the tools has been improved notably, and the tools are expected to improve software development productivity. There are some generation tools, such as Amazon's CodeWhisperer and GitHub's Copilot, and their performances are considered different. To help developers select a tool, Yetiştiren et al. [4] analyzed the performance as the correctness of several tools. Subsequently, the result of the analysis could be utilized as the selection guideline.

Assume that a new code generation tool is released. Although the new tool has the potential to improve the productivity of the development, the tool might also have the risk of declining productivity. To avoid the risk, an analysis of the performance is needed before using the tool. Based on the analysis, developers can decide whether the new tool is better to be used or not. However, such analysis is often time-consuming and, therefore, challenging for developers to perform such analysis whenever a new tool is released.

Instead of such analysis, we propose to evaluate code generation tools during software development (i.e., online optimization) based on the Bandit algorithm (BA). Our significant contributions are twofold. One of them is that we clarified the applicability of BA to the selection of code-generation tools. The other one is we evaluated the performance of BA for the selection.

## 2 Online Evaluation of Code Generation Tools

### 2.1 Bandit Algorithms

BAs are proposed to solve multi-armed bandit problems. Those types of problems are often explained through an analogy with slot machines. Assume that a player has 100 coins to bet on several slot machines, and the player wants to maximize their reward. Instead of selecting only one slot machine and betting all 100 coins, BA suggests that a player bet only one coin on each slot machine. By calculating the average reward of each machine after each betting, the player can then recognize which slot machine is the best (i.e., the highest average reward). The derivation of the problem is that each slot machine has an arm, and the arm is compared to a bandit

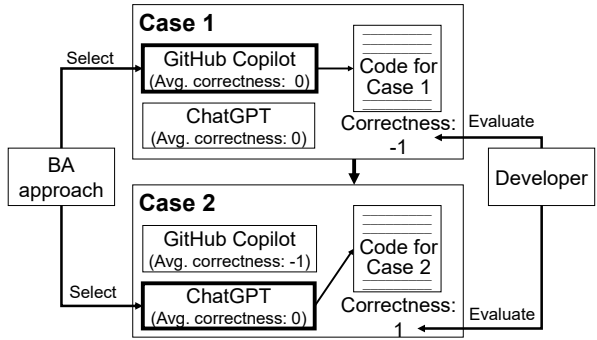

Figure 1: Procedure of tool selection based on BA approach

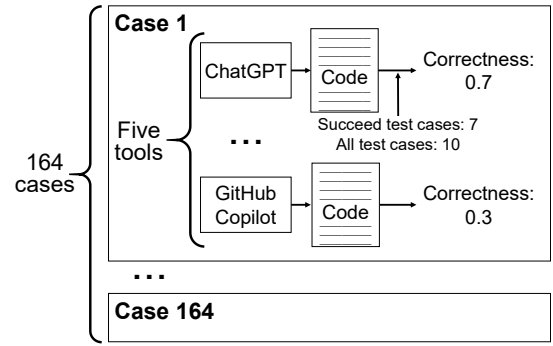

Figure 2: dataset used in the analysis

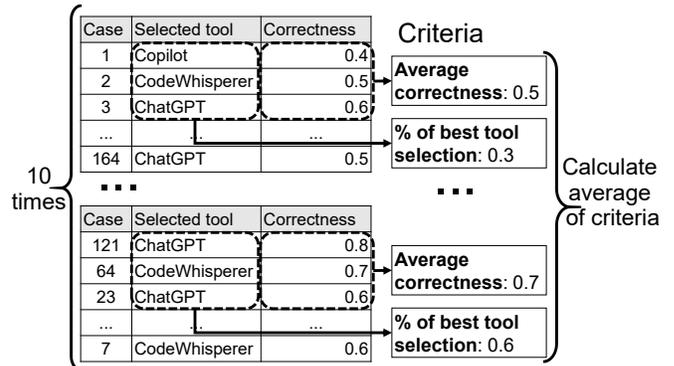

Figure3: Calculation of evaluation criteria

who steals money from players. The multi-armed bandit problem seeks sequentially the best candidates (they are referred to as arms) whose expected rewards are unknown to maximize total rewards.

The epsilon-greedy algorithm is a simple BA method. It chooses a random arm with probability epsilon. It selects an arm with the highest average reward among arms with the probability 1 - ε (0 ≤ ε ≤ 1). When the value of ε is 0, arms are always selected based on the average reward of each arm. In contrast, when the value of ε is 1, arms are always selected randomly.

## 2.2 Proposed Approach

As an analysis approach, we propose to evaluate code generation tools during coding. The procedure of our approach based on BA is the following:

1. A tool is selected randomly.
2. The developer examines the result correctness of the tool (and utilizes the result).
3. The average correctness of each tool is calculated.
4. If the average correctness of the selected tool is low, another tool is selected.
5. Back to Step 2.

Step 2 is indispensable to use generation tools, regardless of our approach, because the generated code is not always correct, even if the best performance tool is used. Based on BA, our approach adds Steps 3 and 4 (i.e., evaluation and selection). Intuitively, our approach considers different generation tools as arms, and developers' evaluation of each tool is the reward.

We show a concrete example of the behavior of our approach. Initially, the average correctness of all tools is set to zero. Hence, for the first case in Fig. 1, GitHub Copilot is selected randomly in step 1. In step 2, a developer evaluates if the generated code is correct, and the correctness is then set to -1 since the developer regards it as "incorrect." Based on the evaluation, we calculate the average correctness for each tool. ChatGPT is selected in step 4 because the average correctness of ChatGPT is higher than that of GitHub Copilot. For the second case, in step 2, the developer regards the generated code "correct." In step 4, based on the evaluation, the correctness is set to 1 since the developer regards it as "correct."

In the domain of software engineering, BAs have been utilized to select methods such as feature selection methods for defect prediction [3] and configuration of clone detection methods [2]. However, no study applies BAs to select code generation tool to the best of our knowledge. The definition of the reward is different from the past studies; therefore, it is unclear whether BAs will work well in selecting code-generation tools.

## 3 Preliminary Analysis

**Dataset**: To evaluate the effectiveness of our approach, we performed a preliminary analysis. In the analysis, we used the dataset published by Yetiştiren et al. [4]. As shown in Fig. 2, the dataset includes 164 cases. In each case, the correctness of code generation tools is evaluated by the number of successful test cases divided by the number of all test cases. The value range of the correctness is [0, 1], and a larger value denotes higher correctness.

Yetiştiren et al. automatically calculated the correctness using test cases from the other dataset [1]. However, such test cases are generally not provided before using the code generation tool, and therefore, developers would manually evaluate correctness (see Step 2 in Section 2.2).

**Code generation tool**: The dataset includes the correctness of five code generation tools. The tools are GitHub Copilot (Version 1.7.4421 and 1.70.8099), Amazon CodeWhisperer (Version Nov.22 and Jan.23), and ChatGPT. In the analysis, BA tried to identify the best generation tool from five candidates of code generation tools. Table 1 shows the average correctness of each tool

Table 1: Average correctness of each tool [4]

| Tool | Copilot Ver.1.7.4421 | Copilot Ver.1.70.8099 | CodeWhisperer Ver.Nov.22 | CodeWhisperer Ver.Jan.23 | ChatGPT |
|---|---|---|---|---|---|
| Average correctness | 0.54 | 0.60 | 0.37 | 0.52 | 0.78 |

Table 2: Average correctness of the proposed approach

| Method | random (baseline) | $\varepsilon = 0.1$ | $\varepsilon = 0.2$ | $\varepsilon = 0.3$ |
|---|---|---|---|---|
| Average correctness | 0.56 | 0.63 | 0.65 | 0.68 |

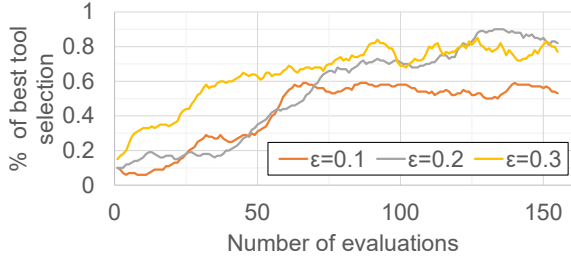

Figure 4: Relationship between the number of evaluations and the percentage of best tool selection

in the 164 cases. As shown in the table, ChatGPT attained the highest correctness value.

**Evaluation criteria**: As an evaluation criterion, we used the average correctness and the percentage of best tool selection. As shown in Fig. 3, the average correctness is BA's average value of tool correctness.

For the analysis, 164 iterations of BA were performed. We analyzed the frequencies of the selected best tool (i.e., ChatGPT), picking up successive ten iterations within the 164 iterations. We call the number of the selection "the percentage of the best tool selection." For instance, as shown in Fig. 3, when ChatGPT is selected once from the first to third iterations, we regarded the percentage as 33% at the third iteration (calculated based on the moving average).

The performance of BA could be affected by the order of the correctness evaluation. Therefore, we randomly sorted 164 cases and evaluated the performance of BA 10 times. After that, we calculated the average of evaluation criteria, as shown in Fig. 3.

**Bandit Algorithm**: For our BA setup, we used ε-greedy (ε = 0.1, 0.2 and 0.3). The performance of BA is affected by the order of cases. Therefore, we randomly sorted the cases and calculated the average of the evaluation criteria, as shown in Fig. 3.

In the analysis, we set the following research questions:

- **RQ1**. How effective is our proposed approach?
- **RQ2**. How often should developers evaluate the candidate tools to identify the best one?

For instance, when BA selects a tool whose correctness is 0.4 two times and selects a tool whose correctness is 0.8 two times, the average correctness is 0.6. RQ1 is set to consider the correctness of BA, and we calculated the average correctness when the tools were selected randomly for the baseline. To answer RQ2, we used the percentage of the best tool selection explained above.

**Result**: Table 2 shows the average correctness of the proposed approach and the baseline. The table shows that our approach is higher than the baseline, and the average correctness was the highest when $\varepsilon = 0.3$. Additionally, compared with Table 1, the average correctness of our approach was still higher than the second one (i.e., GitHub Copilot Ver.1.70.8099). Therefore, the answer to RQ1 is that our approach attains the second-best performance.

Figure 4 shows the relationship between the number of evaluations and the percentage of best tool selection. When the number of evaluations was more than 75, the percentage of best tool selection was higher than 0.7 in most cases. Therefore, the answer to RQ2 is that 75 evaluations are needed to select the best tool with high probability.

## 4  Conclusion

Instead of analyzing the performance of code-generation tools before applying them, we propose a new approach to selecting the best code-generation tools during software development. The proposed approach is based on the Bandit algorithm. In the preliminary analysis, we tried to select the best tool from five candidates of generation tools, such as ChatGPT, using the proposed approach. During the evaluation, our approach based on BA showed the second-best performance, even when BA was not considered for tool selection and without prior analysis.

## ACKNOWLEDGMENTS

This research is partially supported by the Japan Society for the Promotion of Science (JSPS) - Grants-in-Aid for Scientific Research (C)  No. 21K11840.